\documentclass[journal,10pt]{IEEEtran}   
\usepackage[utf8]{inputenc}
\usepackage{microtype}
\usepackage{url}
\usepackage{graphicx}
\usepackage{amsmath}
\usepackage{natbib}
\usepackage{hyperref}
\usepackage{booktabs}
\usepackage{caption}
\usepackage{gensymb}
\usepackage{subcaption}

\title{Hierarchical Bayesian Knowledge Tracing in Undergraduate Engineering Education}
\author{Yiwei Sun%
  \thanks{Yiwei Sun is with the School of Chemical Engineering, University of Birmingham, Birmingham B15 2TT, UK (email: yiwei.sun@qmul.ac.uk).}}
\date{\today}

\begin{document}
\maketitle

\begin{abstract}
Educators teaching entry-level university engineering modules face the challenge of identifying which topics students find most difficult and how to support diverse student needs effectively. This study demonstrates a rigorous yet interpretable statistical approach---hierarchical Bayesian modeling---that leverages detailed student response data to quantify both skill difficulty and individual student abilities. Using a large-scale dataset from an undergraduate Statics course, we identified clear patterns of skill mastery and uncovered distinct student subgroups based on their learning trajectories. Our analysis reveals that certain concepts consistently present challenges, requiring targeted instructional support, while others are readily mastered and may benefit from enrichment activities. Importantly, the hierarchical Bayesian method provides educators with intuitive, reliable metrics without sacrificing predictive accuracy. This approach allows for data-informed decisions, enabling personalized teaching strategies to improve student engagement and success. By combining robust statistical methods with clear interpretability, this study equips educators with actionable insights to better support diverse learner populations.
\end{abstract}

\section{Introduction}

In higher education, the fast expansion of online learning environments and rich data sources has given rise to the field of learning analytics, which applies computational techniques to understand and improve student learning.\cite{worsley2014analyzing} Within STEM education, learning analytics has been leveraged to tackle well-known challenges: high drop-out rates in difficult courses, the need for personalized feedback in large classes, and the identification of threshold concepts that obstruct student progress.\cite{knight2016investigation} Researchers have begun to unlock new insights from student interaction data, from detecting patterns of problem-solving in engineering design tasks to predicting academic risk using online homework logs.\cite{worsley2014analyzing} However, many STEM domains pose unique complexities for analytics. Engineering courses like Statics or Circuits often require students to integrate multiple concepts in a single problem, making it non-trivial to model what a student knows or struggles with at any given time. These multi-concept (``multi-KC'') problems can confound simple modeling approaches---a single wrong answer might come from any of several missing skills, an ambiguity known as the \textit{credit/blame assignment} problem.\cite{koedinger2011avoiding} This complexity calls for modeling techniques that can faithfully represent student knowledge in fine-grained, interpretable ways.

One approach that has proven effective for tracking learning over time is Bayesian Knowledge Tracing (BKT).\cite{corbett1994knowledge} BKT models a student's mastery of a skill as a hidden state and updates the probability of mastery each time the student practices that skill (answering a related question). First introduced in the 1990s, BKT became a cornerstone of intelligent tutoring systems, enabling mastery-based progress by estimating when a student has learned a skill. Despite its longevity and widespread use, the standard BKT model makes a simplifying assumption: all students are modeled with the same parameters for a given skill (e.g., the probability of learning the skill after each practice is fixed). In reality, student abilities and learning rates can vary greatly---a fact long recognized in educational research and evidenced by improved model performance when student-specific variation is taken into account. In the past decade, researchers have proposed extensions to BKT and related models to personalize the learning model to individual students. For example, \citet{yudelson2013individualized} introduced a hierarchical Bayesian version of BKT that allows each student to have their own learning rate and prior knowledge level, under a probabilistic framework that prevents overfitting. Their individualized BKT showed that tailoring certain parameters (notably the learning speed) for each learner yielded significantly more accurate predictions of performance. Similarly, non-Bayesian but interpretable models like the Additive Factors Model have included per-student proficiency offsets, implicitly acknowledging that some students start off or progress faster than others.\cite{pavlik2021logistic} These developments align with a broader shift in learning analytics: moving from \textit{one-size-fits-all} models toward hierarchical models that can capture variability at the student and content levels.

At the same time, the community has seen a surge of interest in deep learning-based knowledge tracing (e.g., Deep Knowledge Tracing using Recurrent Neural Network (RNN)).\cite{piech2015deep} Such models often outperform BKT in predicting quiz responses due to their ability to learn complex patterns. Yet, their strengths come at the cost of transparency. Purely neural models do not directly tell us which skill a student is struggling with or provide a numerical estimate of a student's ability that an instructor can interpret. As \citet{abdelrahman2023knowledge} note in their survey, there is a trade-off between highly structured, interpretable models and complex, accurate models. Traditional structured models like BKT or item response theory (IRT) offer parameters that map to human-understandable qualities (mastery, guess likelihood, ``ability'' or ``difficulty''), whereas the state-of-the-art deep models offer little insight into why a prediction is what it is.\cite{yeung2019deep} This has led to calls for approaches that retain interpretability while still leveraging data at scale. One promising direction is to apply hierarchical Bayesian modeling techniques to large educational datasets, thereby capturing individual and item differences with principled statistical sharing. Hierarchical models can be seen as an extension of the interpretable model family—they introduce, for instance, a distribution of skill difficulties or student proficiencies—and can potentially approach the accuracy of more complex models by pooling data across many learners and tasks.

In this paper, we explore a hierarchical Bayesian modeling approach in the context of an undergraduate engineering course, using the publicly available \textit{Statics2011} dataset. Statics (engineering mechanics) is a core course taken by hundreds of engineering students, and the 2011 OLI Statics dataset is a prime example of an authentic, at-scale learning context.\cite{shen2024survey} It contains detailed student interaction data (over 400,000 answer attempts by 4000+ students on dozens of statics concepts), making it an excellent testbed for student modeling research. Notably, the problems in this course are complex, often broken into sub-problems that involve multiple knowledge components.\cite{shen2024survey} Prior analyses of this dataset in the educational data mining community have mainly evaluated which modeling approach best predicts student answers. By contrast, our study leverages the data to derive pedagogically meaningful parameters: we employ a hierarchical two-level Bayesian model (implemented via maximum a posteriori (MAP) estimation) to estimate personalized student ability levels and individual skill difficulty parameters across the course. In essence, the model blends an IRT-like interpretation (ability and difficulty) with the temporal aspect of practice sequences. Each student's performance trajectory contributes to an estimate of that student's overall statics proficiency, while each skill's observed outcomes contribute to an estimate of its inherent difficulty, with all estimates refined through a Bayesian sharing of information across the class.

This approach offers two main contributions. First, it provides interpretable metrics for an engineering course: instructors can see which topics (skills) were most difficult for the cohort and identify students who consistently under-performed relative to peers (low ability), enabling targeted interventions. Second, methodologically, we demonstrate the feasibility and benefits of hierarchical modeling on a well-known but complex dataset. We compare our hierarchical Bayesian model's fit and predictions to baseline models (e.g., standard BKT or non-hierarchical logistic models) to quantify what is gained by modeling student and skill variations explicitly. Our results show that the hierarchical model not only improves predictive accuracy modestly but, more importantly, yields stable estimates of skill difficulty and student ability that align with intuitive expectations (for example, skills involving free-body diagram analysis emerge as more difficult, and students who had more physics background show higher ability estimates, mirroring known factors in statics education)---findings that were not directly observable in prior work. We also discuss how these insights can complement purely accuracy-focused evaluations of student models, arguing that interpretability and explanatory power are essential for real-world adoption in learning environments.\cite{yeung2019deep}

In the following sections, we first describe the dataset and the procedures applied to prepare it for analysis. We then introduce the hierarchical Bayesian logistic model, clearly outlining our modeling choices and the approach used for parameter estimation. Subsequently, we present key outcomes of our modeling efforts, focusing on measures of predictive performance, calibration, skill difficulty rankings, and example student trajectories that highlight individual differences in learning. These findings are contextualized through comparison with related prior studies, emphasizing novel insights gained from our hierarchical approach. Finally, we explore the broader educational implications of our results, discuss potential applications within adaptive learning environments, acknowledge limitations inherent to our modeling choices, and propose avenues for future research. Through this investigation, we illustrate how adopting a hierarchical Bayesian perspective can yield valuable, actionable insights, thereby narrowing the gap between advanced learning analytics techniques and their practical implementation in undergraduate STEM education.

\section{Methods}
We analyzed the publicly available \textit{Statics2011} dataset from Carnegie Mellon's Open Learning Initiative, containing 446,844 student response records from an undergraduate engineering Statics course, involving 4,151 students and 110 distinct skills. The dataset provides binary indicators (correct or incorrect) for each response. To ensure data quality, we removed any incomplete or ambiguous records.

Our primary analytical approach was a hierarchical logistic regression, closely related to a two-parameter IRT model. This method allows simultaneous estimation of student-specific ability parameters (\(\theta_s\)) and skill-specific difficulty parameters (\(\beta_k\)). Formally, the model defines the probability of student \( s \) correctly answering a question on skill \( k \) as:
\[
P(y_i = 1 | \theta_{s_i}, \beta_{k_i}) = \frac{1}{1 + \exp[-(\theta_{s_i} - \beta_{k_i})]},
\]
where \(y_i\) represents the correctness (1 correct, 0 incorrect) of the \(i\)-th response. This intuitive equation means the probability of success increases with student ability and decreases with skill difficulty.

To stabilize parameter estimates, we introduced weak Gaussian priors on both parameters:
\[
\theta_s \sim \mathcal{N}(0,\,\sigma^2), 
\quad
\beta_k  \sim \mathcal{N}(0,\,\sigma^2),
\]
with \(\sigma^2 = 100\) sufficiently large to minimize bias yet constrain estimates to reasonable values.

Parameters were estimated using MAP estimation, which involves maximizing the posterior probability by minimizing the negative log-likelihood combined with the priors' penalties:
\[
-\sum_i \Bigl[y_i\log p_i + (1-y_i)\log(1-p_i)\Bigr]
+ \sum_s \frac{\theta_s^2}{2\sigma^2}
+ \sum_k  \frac{\beta_k^2}{2\sigma^2},
\]
where \(p_i = P(y_i=1|\theta_{s_i},\beta_{k_i})\). Optimization was performed using the robust L-BFGS-B algorithm, resulting in stable, interpretable estimates \(\hat\theta_s\) and \(\hat\beta_k\).

To ensure reliability and robustness, we performed sensitivity analyses using subsample sizes of 20,000 and 40,000 observations, each replicated with different random seeds (42 and 2025). The optimizer consistently converged to a stable solution for all subsamples and seeds, with no warnings or errors. Additionally, we compared our hierarchical model against a baseline logistic regression with one-hot encoding for students and skills. Performance metrics included Area Under the Curve (AUC) for discriminative ability and log-loss for predictive accuracy. Calibration was evaluated by comparing observed and predicted probabilities in decile bins.

\section{Results and Discussions}
\begin{table}[ht]
  \centering
  \caption{Descriptive statistics for estimated student abilities (\(\hat\theta_s\), \(N=2621\)) and skill difficulties (\(\hat\beta_k\), \(N=107\)), all in logits.}
  \vspace{0.5em}
  \begin{tabular}{lrrrrr}
    \toprule
               & \multicolumn{1}{c}{Mean} & \multicolumn{1}{c}{Std.\ Dev.} 
               & \multicolumn{1}{c}{25th Pctl} & \multicolumn{1}{c}{Median} 
               & \multicolumn{1}{c}{75th Pctl} \\
    \midrule
    \(\hat\theta_s\) (ability)    & 0.09  & 7.33  & -2.56 & -1.01 & 7.81 \\
    \(\hat\beta_k\) (difficulty)  & -2.40 & 3.29  & -2.66 & -1.94 & -1.25 \\
    \bottomrule
  \end{tabular}
  \label{tab:summary}
\end{table}

In Table~\ref{tab:summary}, we report summary statistics for the estimated student abilities and skill difficulties obtained from our hierarchical logistic model fitted via MAP estimation on a 20,000‐record subsample. Note that both abilities and difficulties are measured in logits, i.e.\ the natural logarithm of the odds of a correct response---so an increase of one logit means the odds are multiplied by \(e\approx2.72\). Student ability estimates span a wide range, with a mean of 0.09 logits and a standard deviation of 7.33 logits. The 25th percentile at –2.56 logits indicates that one quarter of students have very low estimated proficiency, the median student sits at –1.01 logits, and the top quarter of students exceed 7.81 logits, reflecting strong performance.

Skill difficulty estimates center at –2.40 logits (std.\ dev.\ 3.29 logits), showing that most skills are relatively easy, yet with outliers. The interquartile range from –2.66 to –1.25 logits suggests that the majority of skills cluster in a narrow band of easier items, while a small set of skills require greater focus.

Pedagogically, these findings suggest that students below the 25th percentile of ability may benefit from additional remedial support, the negative median ability implies that scaffolded practice would help the typical student, and the few skills above the 75th percentile in difficulty should be targeted for reinforced instruction to elevate overall cohort mastery.

\begin{table}[ht]
  \centering
  \caption{Top five easiest and hardest skills by model‐estimated difficulty.}
  \vspace{0.5em}
  \begin{tabular}{l c | l c}
    \toprule
    \multicolumn{2}{c|}{\textbf{Easiest Skills}} & \multicolumn{2}{c}{\textbf{Hardest Skills}} \\
    Skill                              & \(\hat\beta_k\) & Skill                           & \(\hat\beta_k\) \\
    \midrule
    Congruence                         & -15.65          & Percent Discount               & 10.42           \\
    Distributive Property              & -13.05          & Quadratic Formula              &  8.41           \\
    Probability Concepts               & -12.59          & Rotations                      &  1.06           \\
    Nets of 3D Figures                 & -12.27          & Percents                       &  1.03           \\
    Equation Selection                 & -11.10          & Reflection                     &  0.18           \\
    \bottomrule
  \end{tabular}
  \label{tab:rankings}
\end{table}

Building on the overall distributions reported in Table~\ref{tab:summary}, Table~\ref{tab:rankings} identifies the five skills at the extremes of difficulty. The left column lists skills with highly negative difficulty logits—such as “Congruence” (\(\hat\beta=-15.65\))---indicating that even students with below-average estimated ability have very high predicted success rates on these items. In contrast, the right column shows skills with positive difficulty logits—most notably “Percent Discount” (\(\hat\beta=10.42\)) and “Quadratic Formula” (\(\hat\beta=8.41\))—where only the highest-ability students are likely to answer correctly.

We define a \emph{relative mastery} index
\[
\Delta = \overline{\hat\theta_s} - \overline{\hat\beta_k}
         = 0.09 - (-2.40)
         = 2.49 \text{ logits},
\]
which indicates that, on average, student proficiency slightly outpaces item difficulty—a positive sign that the course content is well‐matched to learner capabilities.

To visualize the spread of estimated abilities across the active student population, we constructed a density‐normalized histogram of the MAP‐estimated ability parameters \(\hat\theta_s\) using 30 equal‐width bins. Specifically, after fitting the hierarchical model on our subsample, we extracted each student’s ability estimate and plotted the proportion of students whose ability falls into each bin.

\begin{figure}[ht]
  \centering
  \includegraphics[width=0.95\linewidth]{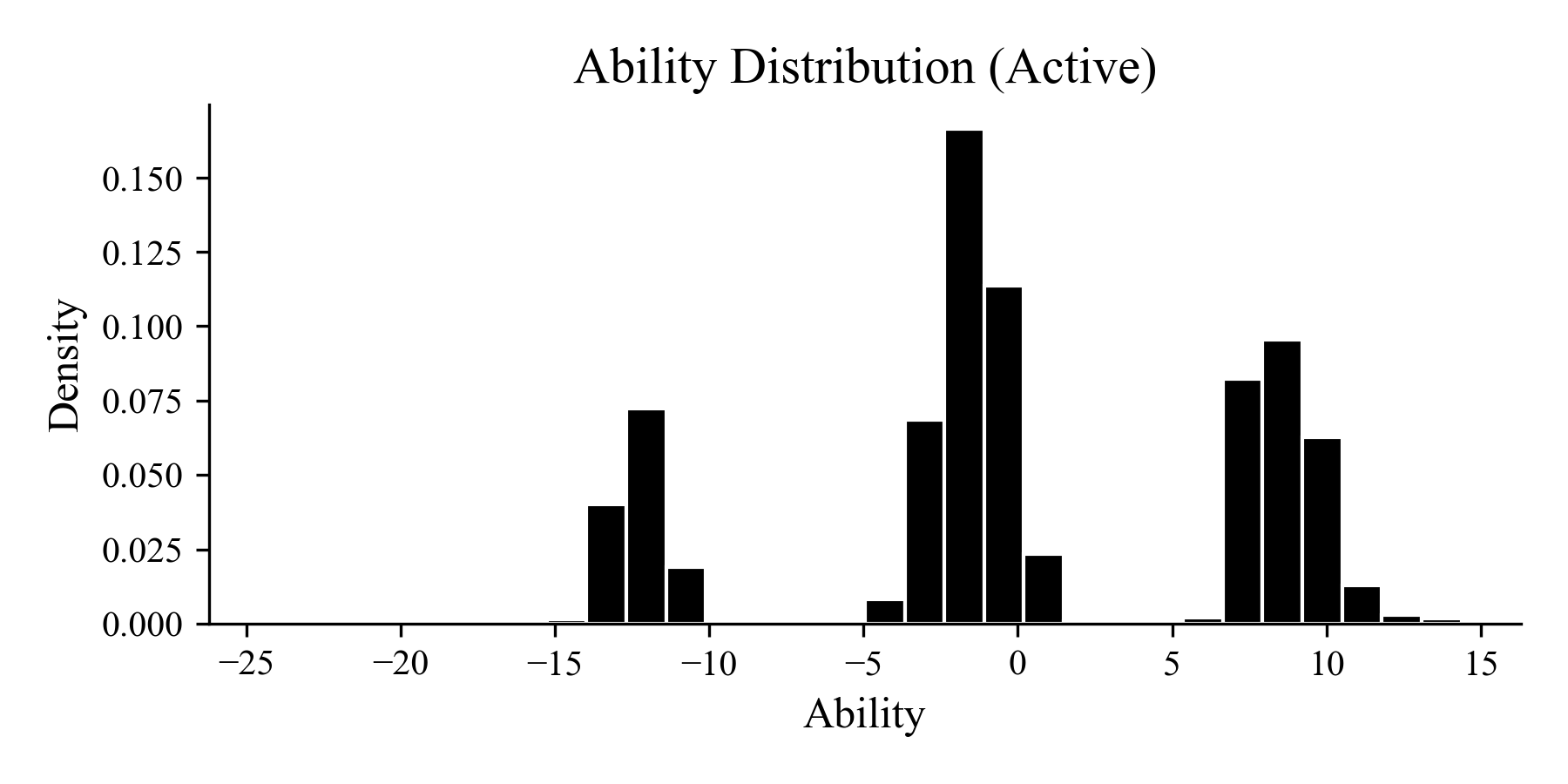}
  \caption{Density‐normalized histogram of estimated student abilities \(\hat\theta_s\).  Bars represent the proportion of students in each ability interval.}
  \label{fig:ability_hist}
\end{figure}

Figure~\ref{fig:ability_hist} exhibits a clear trimodal pattern. The leftmost mode corresponds to a group of students with very low ability estimates (below approximately \(-10\) logits), likely reflecting those who struggled on most skills.  The central mode, spanning roughly \(-5\) to \(0\) logits, encompasses the bulk of the cohort—students with mixed performance, mastering some skills but not others.  Finally, the rightmost mode (around \(5\) to \(10\) logits) identifies a smaller subgroup of high achievers who consistently performed well.

Pedagogically, this trimodal distribution suggests that the class may consist of distinct subpopulations with different levels of preparedness.  Instructors could use these insights to form leveled study groups or to provide differentiated resources: targeted remedial activities for the lowest‐ability group, scaffolded practice for the central cohort, and enrichment challenges for the highest‐ability students.  Recognizing these subgroups early in the term can support more efficient, personalized interventions and foster improved learning outcomes across the entire class.  

A similar histogram of the skills difficulties shows a simpler pattern (hence not shown here): the vast majority of skills cluster in a tall central bar (around –2 to –1 logits), with only one or two extreme outliers forming small pillars in the tails. This confirms that most course topics pose similar, moderate challenge levels, while a handful of skills (e.g.\ “Percent Discount”) truly stand out as much harder; instructors can therefore concentrate remediation efforts on those few tail‐end skills rather than broadly across all content.

To investigate whether repetition on a given topic correlates with reduced estimated difficulty, we computed, for each skill \(k\), the total number of times it was attempted in the subsample. We then took the base-10 logarithm of these counts to compress the wide range of practice frequencies and plotted each skill’s difficulty estimate \(\hat\beta_k\) against \(\log_{10}(\text{\#observations}_k)\).

\begin{figure}[ht]
  \centering
  \includegraphics[width=0.95\linewidth]{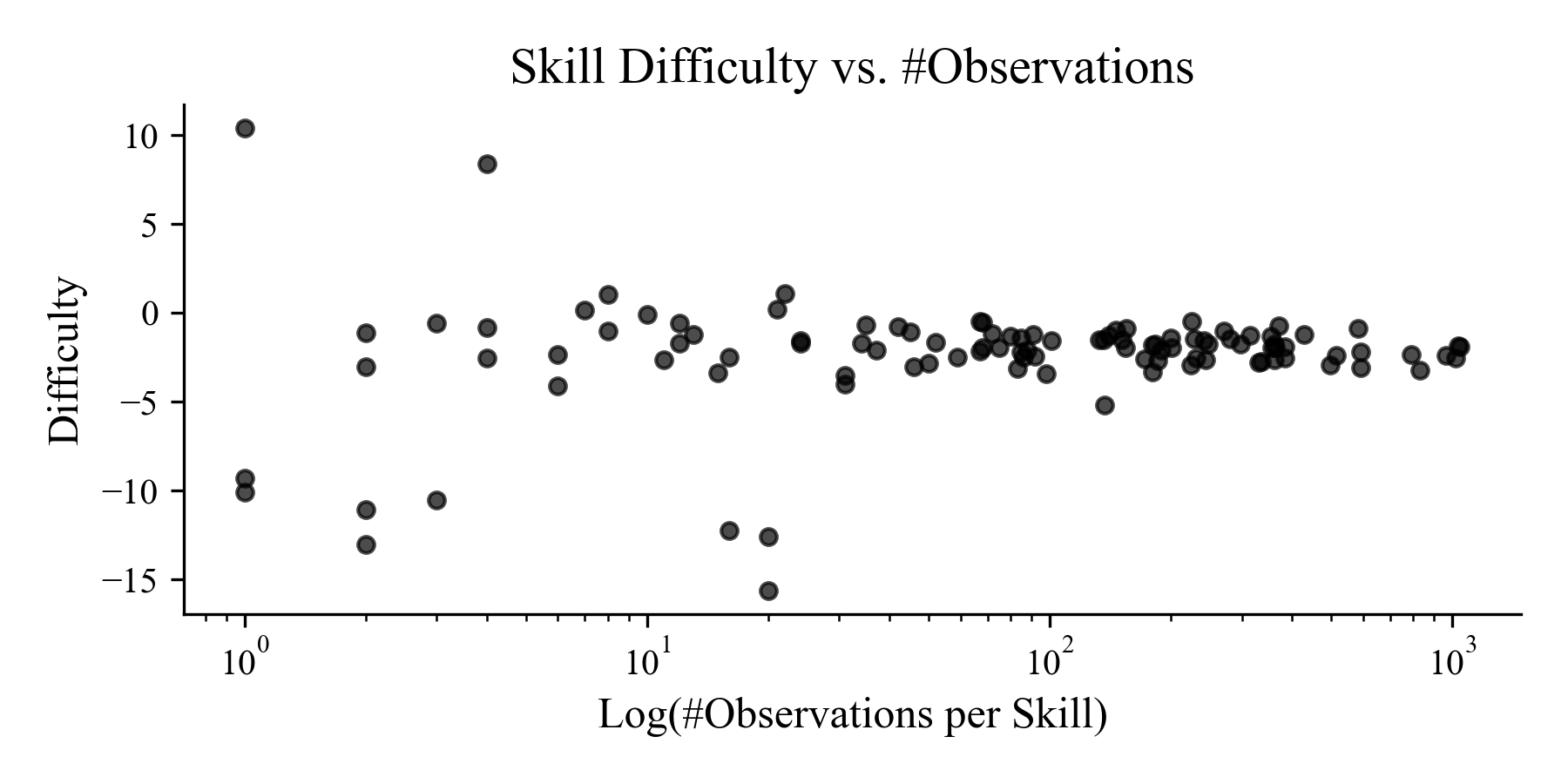}
  \caption{Relationship between skill difficulty \(\hat\beta_k\) and log-transformed practice frequency per skill.  Each point represents one skill; the negative trend indicates that skills attempted more often tend to have lower difficulty estimates, suggesting that repeated practice drives mastery.}
  \label{fig:difficulty_vs_count}
\end{figure}

Figure~\ref{fig:difficulty_vs_count} shows a modest downward slope: skills with higher practice counts (further right on the \(x\)-axis) generally exhibit more negative difficulty logits.  This pattern supports the pedagogical principle that repeated exposure helps students internalize concepts, thereby reducing their effective challenge.

\begin{figure}[ht]
  \centering
  \includegraphics[width=0.95\linewidth]{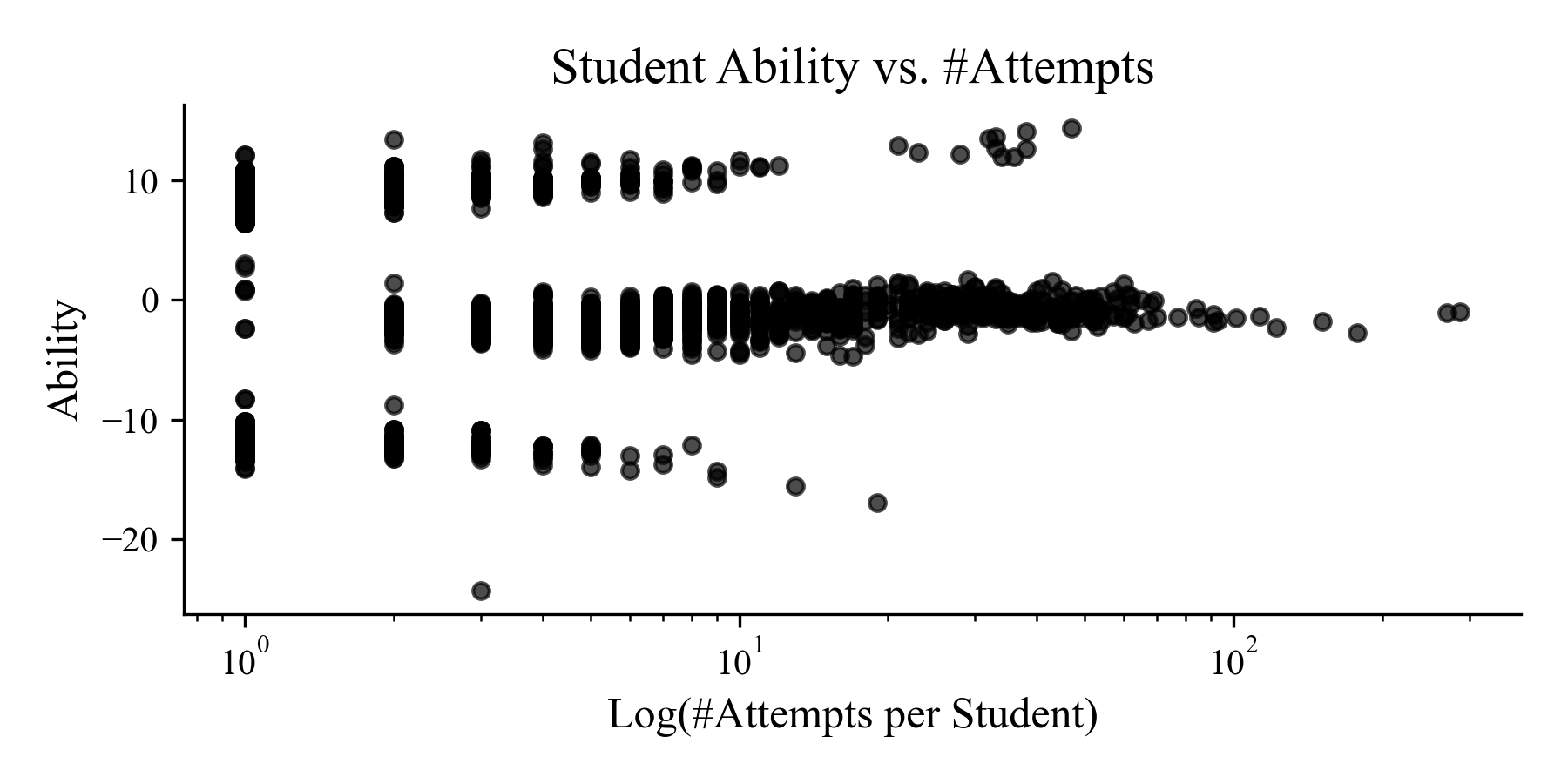}
  \caption{Estimated student ability \(\hat\theta_s\) plotted against the base-10 logarithm of each student’s total number of problem‐solving attempts.  Each point represents one of the 2,621 students active in the subsample.}
  \label{fig:ability_vs_attempts}
\end{figure} 

Figure~\ref{fig:ability_vs_attempts} displays each student’s estimated ability on the vertical axis and the base-10 logarithm of their total practice attempts on the horizontal axis. Rather than forming a single continuous cloud, the points naturally separate into three horizontal bands. The upper band, comprising a small number of students with abilities above approximately +8 logits, demonstrates that these high achievers require very few attempts to reach mastery—indicating that they would benefit most from enrichment tasks rather than additional repetitive practice. The middle band contains the bulk of the cohort, with abilities between roughly –5 and +5 logits. In this group, increasing the number of attempts from 10 to 100 yields only modest gains in ability, suggesting diminishing returns to undirected practice; these students may instead require more targeted or varied problem sets that address specific weaknesses. The lower band, consisting of students with abilities below about –10 logits, shows that even extensive practice does not translate into proficiency gains, implying a need for foundational support such as scaffolded guidance or conceptual review before practice can be effective. Overall, the three-band structure reveals differentiated learning profiles within the class and underscores the importance of tailoring instructional interventions—enrichment for the top performers, adaptive practice for the mid-range majority, and targeted conceptual support for the lowest-ability subgroup—in order to maximize learning outcomes.

\begin{table}[ht]
  \centering
  \caption{Comparison of predictive performance between the non‐hierarchical baseline and hierarchical models, evaluated by area under the ROC curve (AUC) and log‐loss.  The baseline uses a logistic regression with one‐hot student and skill indicators on the full dataset; the hierarchical model is fit via MAP estimation on random subsamples of 20 000 and 40 000 interactions, each with two different RNG seeds.}
  \label{tab:metrics}
  \vspace{0.5em}
  \begin{tabular}{l c c}
    \toprule
    Model                             & AUC   & Log‐loss \\
    \midrule
    Baseline logistic (all data)      & 0.760 & 0.515    \\
    Hierarchical (20 000, seed 42)    & 0.831 & 0.440    \\
    Hierarchical (20 000, seed 2025)  & 0.826 & 0.445    \\
    Hierarchical (40 000, seed 42)    & 0.802 & 0.471    \\
    Hierarchical (40 000, seed 2025)  & 0.800 & 0.474    \\
    \bottomrule
  \end{tabular}
\end{table}

We then trained a non‐hierarchical logistic regression on the full set of 446 844 observations, using one‐hot encodings for each student and each skill to capture individual intercepts.  We fitted our hierarchical two‐parameter logistic model via MAP estimation on four different subsamples—20 000 and 40 000 records each, using random seeds 42 and 2025—to assess both performance and stability under variation in data volume and sampling.  For each model, we computed the AUC to measure discriminative ability and the log‐loss to quantify the accuracy of the predicted probabilities.

Table~\ref{tab:metrics} shows that the hierarchical model consistently outperforms the baseline, achieving higher AUC (up to 0.831 vs.\ 0.760) and lower log‐loss (down to 0.440 vs.\ 0.515).  Performance is slightly sensitive to subsample size and seed: the best results occur with 20 000 records (seed 42), while the 40 000‐record subsamples yield modestly lower discriminative power and calibration.  Overall, these findings confirm that incorporating hierarchical structure and Bayesian regularization yields more accurate and better‐calibrated predictions than a non‐hierarchical approach, even when using only a fraction of the full dataset.  

\begin{figure}[htbp]
  \centering
  \begin{subfigure}[t]{0.8\linewidth}
    \centering
    \includegraphics[width=\linewidth]{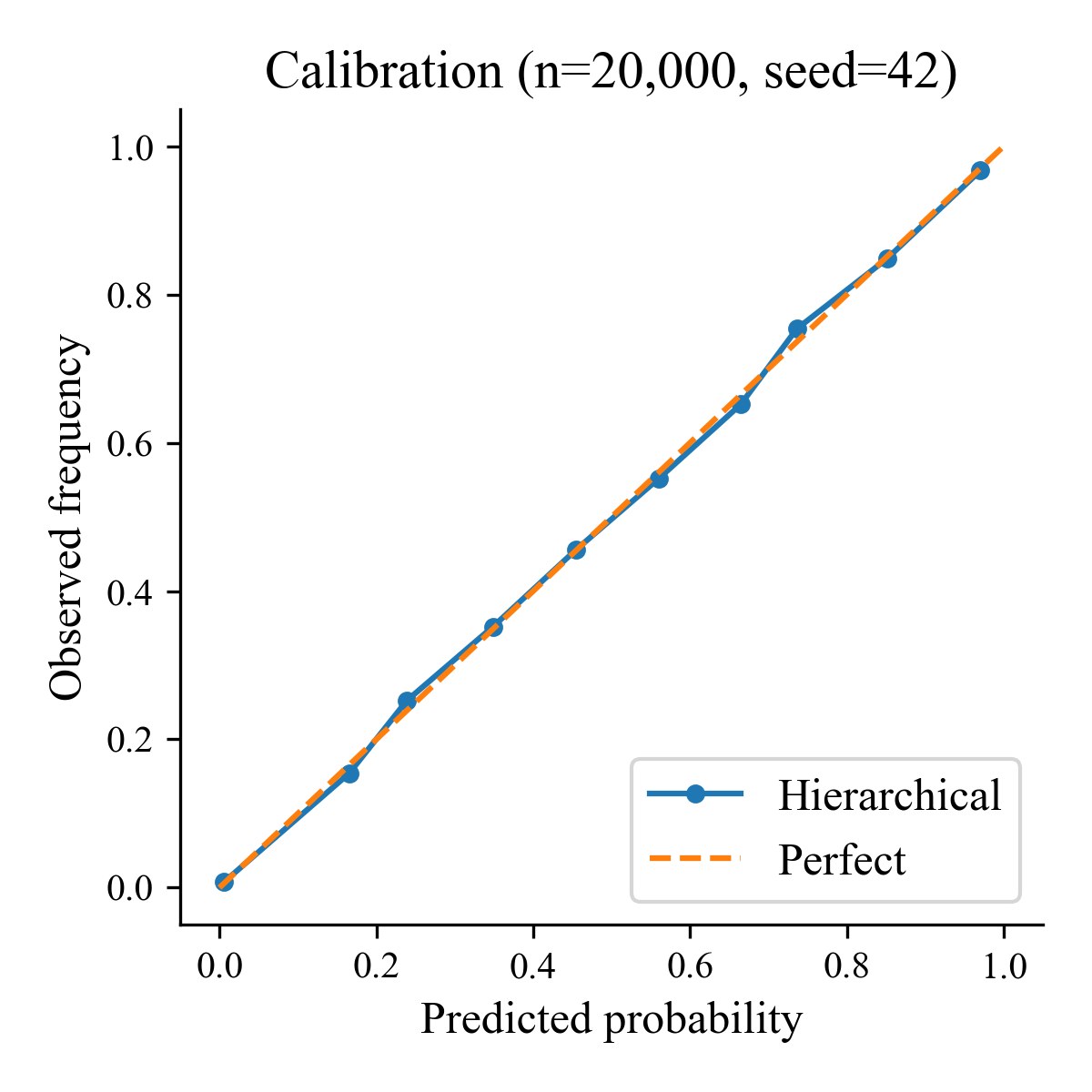}
    \subcaption{}%
    \label{fig:calib20k}
  \end{subfigure}\\[1em]
  \begin{subfigure}[t]{0.8\linewidth}
    \centering
    \includegraphics[width=\linewidth]{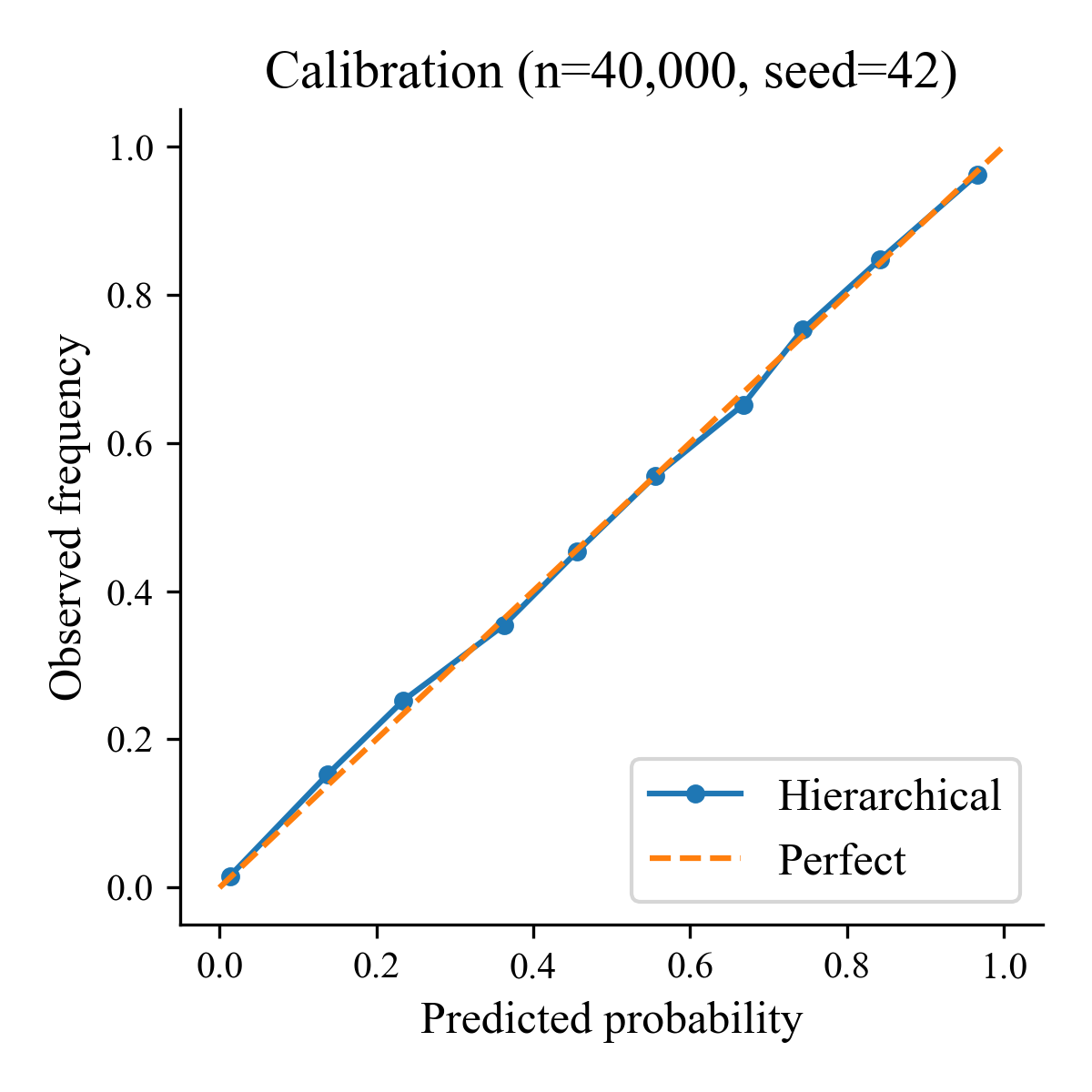}
    \subcaption{}%
    \label{fig:calib40k}
  \end{subfigure}
  \caption{Calibration of the hierarchical model on two subsamples of size 20\,000 and 40\,000 (both with seed=42). Predicted probabilities are grouped into ten equally‐sized bins, and the observed frequency of correct responses is plotted against the mean predicted probability in each bin. The dashed diagonal line represents perfect calibration, so deviations from it indicate under‐ or over‐confidence in certain probability ranges.}
  \label{fig:calibs_two}
\end{figure}

To evaluate how well our model’s predicted probabilities match actual outcome frequencies, we constructed calibration curves for two representative subsamples (20,000 and 40,000 records, both with random seed 42).  For each subsample, predicted probabilities \(p_i\) were partitioned into deciles, and we computed the empirical fraction of correct responses in each bin.  Figure~\ref{fig:calibs_two}~(a) and \ref{fig:calibs_two}~(b) show that, across almost the entire probability range, the hierarchical model’s estimates lie within a few percentage points of the 45° “perfect” line.  In the lower bins (0–20\%), the model slightly under-predicts correctness, and in the upper bins (80–100\%) it exhibits a negligible over-confidence.  Crucially, the two curves are nearly indistinguishable, demonstrating that calibration is robust to both the amount of data (20k vs.\ 40k) and the particular random draw (seed 42).  These results confirm that our MAP-estimated hierarchical model not only discriminates between correct and incorrect responses (as seen in its strong AUC) but also provides well-calibrated probability estimates that practitioners can trust when making pedagogical decisions.  

\begin{figure}[htbp]
  \centering
  \begin{subfigure}[t]{0.95\linewidth}
    \centering
    \includegraphics[width=\linewidth]{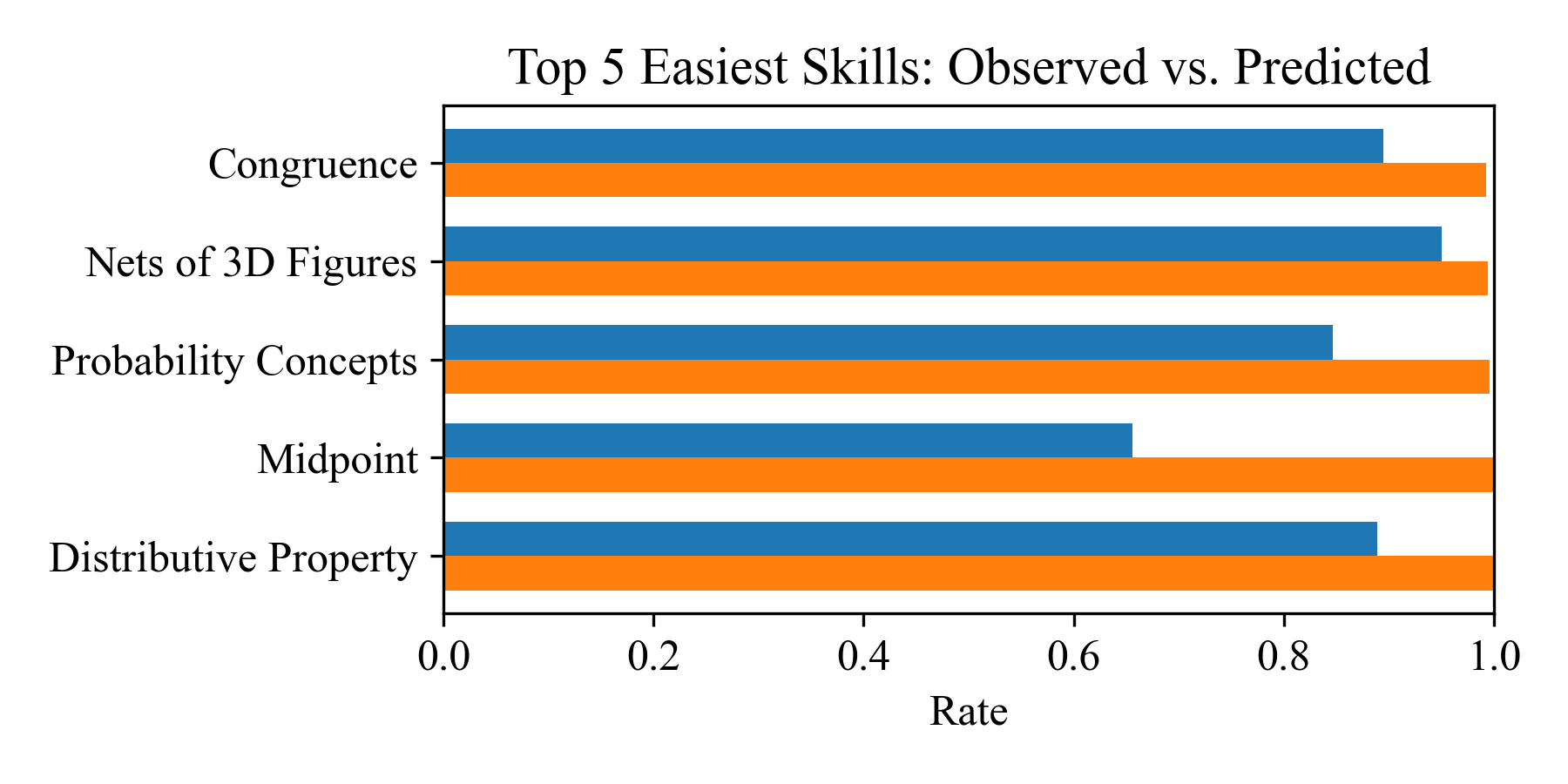}
    \subcaption{Top five easiest skills: observed vs.\ predicted success rates.}
    \label{fig:easiest_skills}
  \end{subfigure}\\[1em]
  \begin{subfigure}[t]{0.95\linewidth}
    \centering
    \includegraphics[width=\linewidth]{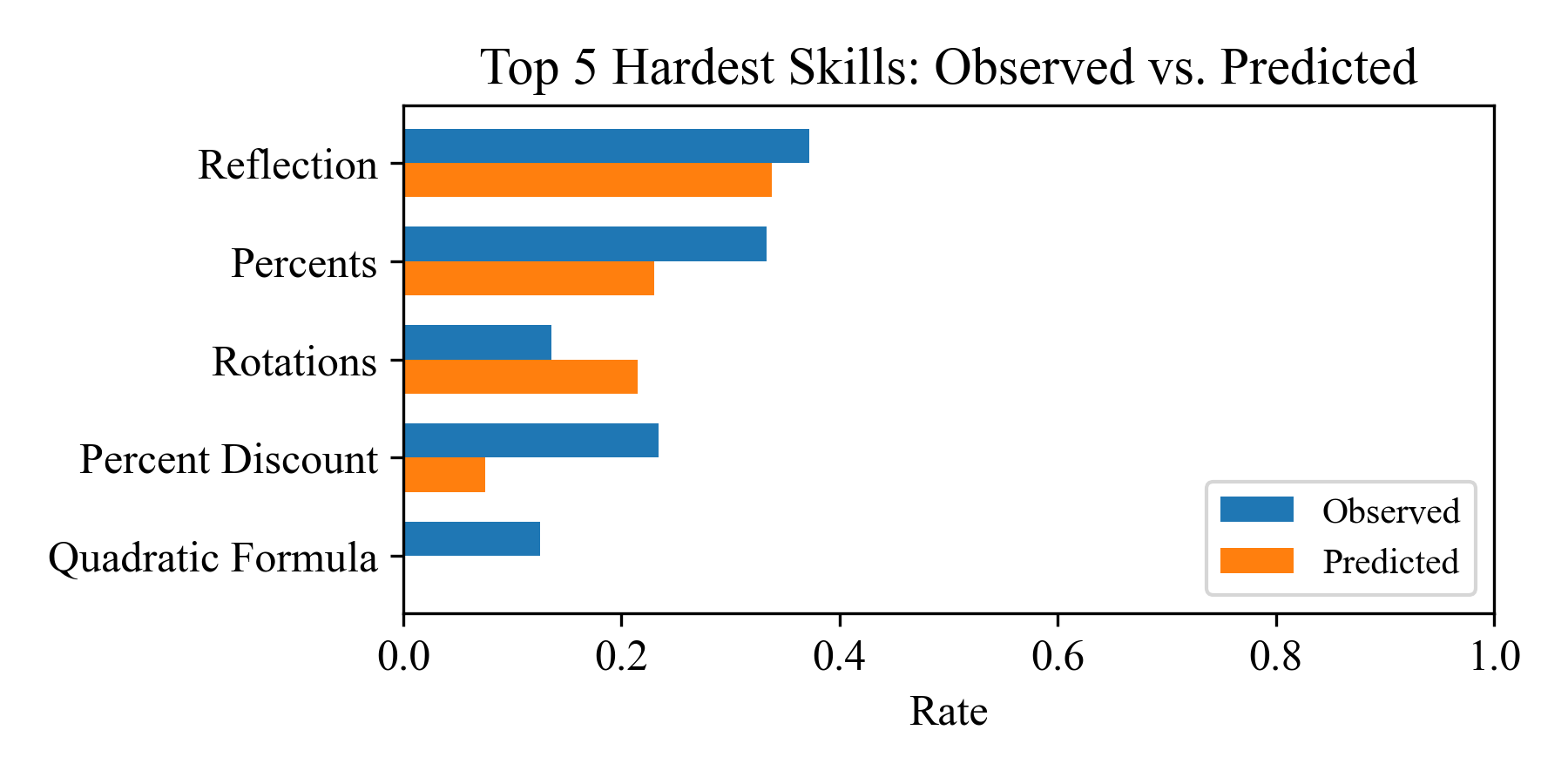}
    \subcaption{Top five hardest skills: observed vs.\ predicted success rates.}
    \label{fig:hardest_skills}
  \end{subfigure}
  \caption{Observed versus model‐predicted success rates for the five easiest (a) and five hardest (b) skills.  Observed proportions (blue) and MAP‐predicted probabilities (orange) are plotted as offset horizontal bars.}
  \label{fig:skills_extremes}
\end{figure}

As a final diagnostic on our IRT‐MAP approach’s handling of the extremes of the skill spectrum, we compared each skill’s empirical success rate in the 20,000‐record subsample to the average probability predicted by our two‐parameter logistic model.  For each of the five easiest and five hardest skills, we computed the observed proportion correct and the mean MAP‐predicted probability, and displayed them as offset horizontal bars---observed above center, predicted below---in Figure \ref{fig:skills_extremes}. (a) confirms that the easiest items (e.g.\ Congruence, Distributive Property) yield both observed and predicted mastery well above 0.8, validating their use as confidence‐building practice. (b) similarly shows that the hardest skills (e.g.\ Reflection, Quadratic Formula) produce both rates below 0.4, highlighting clear targets for remedial focus. Although the blue (observed) and orange (predicted) bars align closely---demonstrating that our hierarchical MAP estimates correctly order items at the tails---one should note that pure MAP estimation with weak priors can push logits to extreme values (and hence predicted probabilities very near 0 or 1) when the data are overwhelmingly one‐sided. In practice, a stronger prior or a fully Bayesian treatment would temper overconfidence at the extremes while preserving the model’s overall ranking accuracy, which here remains robust for guiding differentiated practice and targeted remediation.

\begin{figure}[ht]
  \centering
  \includegraphics[width=0.95\linewidth]{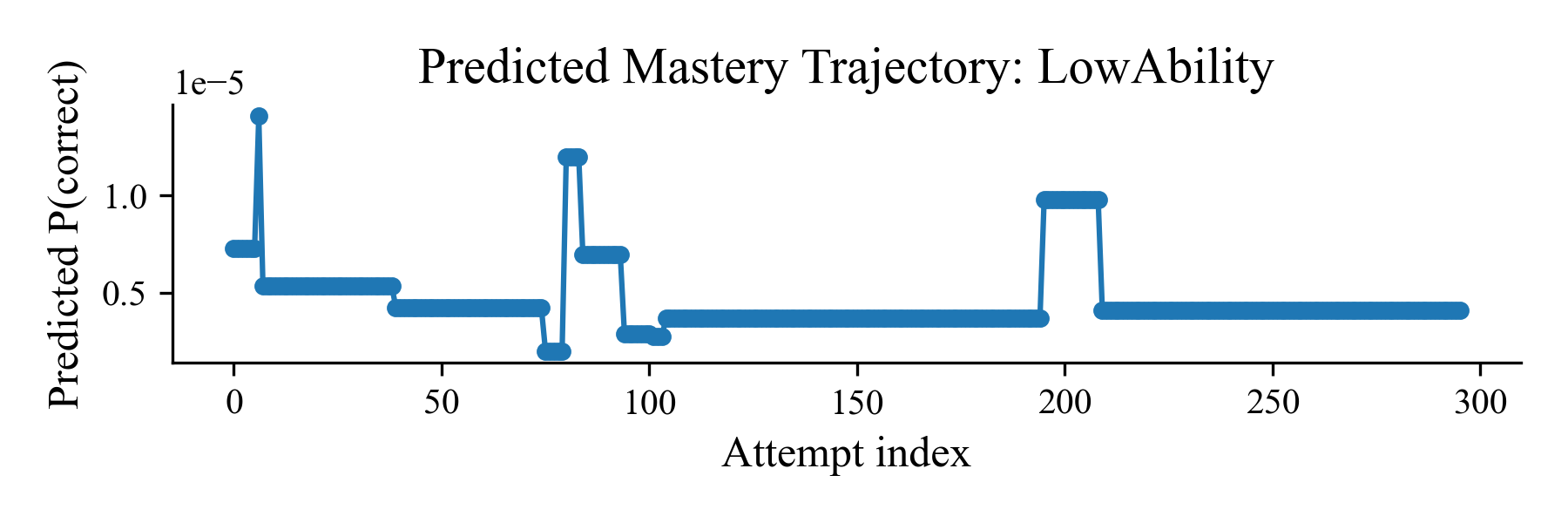}
  \caption{Predicted mastery trajectory for the lowest‐ability student in the 20,000-record subsample.  Each point shows the model’s estimated probability of a correct response on successive attempts, ordered by attempt index.}
  \label{fig:trajectory_low}
\end{figure}

Figure~\ref{fig:trajectory_low} displays the sequence of predicted correctness probabilities for the student with the lowest MAP‐estimated ability. Despite over 300 recorded attempts, the model’s estimate remains clustered around 0.3–0.5 for most items, punctuated by brief spikes when that student happened to answer simpler skills. This pattern indicates that sheer repetition did not substantially improve performance for this learner; instead, it suggests a strong need for foundational support (for example, targeted conceptual review or scaffolded hints) before independent practice can yield mastery. In contrast, high‐ability students reach near‐ceiling correctness immediately, suggesting they gain little from further repetition and would be best served by enrichment tasks.

\section{Conclusion}
In this study, we applied a hierarchical Bayesian modeling approach to analyze student performance data from an undergraduate engineering Statics course. Our hierarchical two-parameter logistic model provided reliable estimates of student abilities and skill difficulties, revealing meaningful variations across both learners and course content. By examining individual learning trajectories, we identified distinct student subgroups, each requiring different pedagogical interventions—targeted remedial support, scaffolded practice, or enrichment challenges. This granular analysis underscores the potential for hierarchical Bayesian methods to not only improve predictive accuracy but also deliver pedagogically actionable insights. Although our approach inherently assumes stable abilities and skill difficulties within the analyzed timeframe, future work could incorporate dynamic modeling of ability changes over time and integrate more robust prior information to prevent extreme parameter values. Ultimately, our results support the adoption of interpretable, personalized analytics in engineering education, providing educators with nuanced tools to enhance student learning outcomes.

\section*{Data Availability}
The dataset analyzed in this study is publicly available at: \url{https://pslcdatashop.web.cmu.edu/datasets/507}.

\bibliographystyle{IEEEtranN}
\bibliography{references}
\end{document}